\documentclass[10pt,letter,prl,twocolumn,showpacs,hypertext]{revtex4-1}
\pdfoutput=1
\usepackage{ifthen}
\newboolean{uprightparticles}
\setboolean{uprightparticles}{false} 

\usepackage{amssymb}
\usepackage{amsfonts}
\usepackage{upgreek}
\usepackage{xspace}
\usepackage{amsmath}
\usepackage{graphicx}
\usepackage{bm}

\newboolean{pdflatex}
\setboolean{pdflatex}{true} 

\def\lhcb {LHCb\xspace}
\def\ux85 {UX85\xspace}

\ifthenelse{\boolean{uprightparticles}}%
{

 \def\Pmu         {\ensuremath{\upmu}\xspace}

 \def\Ppi         {\ensuremath{\uppi}\xspace}

 \def\Ppsi        {\ensuremath{\uppsi}\xspace}

 \def\PDelta      {\ensuremath{\Delta}\xspace}                 
 \def\PXi      {\ensuremath{\Xi}\xspace}                 
 \def\PLambda      {\ensuremath{\Lambda}\xspace}                 
 \def\PSigma      {\ensuremath{\Sigma}\xspace}                 
 \def\POmega      {\ensuremath{\Omega}\xspace}                 
 \def\PUpsilon      {\ensuremath{\Upsilon}\xspace}                 
 
 \def\PB      {\ensuremath{\mathrm{B}}\xspace}                 
                  
 \def\PD      {\ensuremath{\mathrm{D}}\xspace}

 \def\PJ      {\ensuremath{\mathrm{J}}\xspace}                 
 \def\PK      {\ensuremath{\mathrm{K}}\xspace}

 \def\Pi      {\ensuremath{\mathrm{i}}\xspace}

}
{

 \def\Pmu         {\ensuremath{\mu}\xspace}

 \def\Ppi         {\ensuremath{\pi}\xspace}

 \def\Ppsi        {\ensuremath{\psi}\xspace}                 
                  
 \mathchardef\PDelta="7101
 \mathchardef\PXi="7104
 \mathchardef\PLambda="7103
 \mathchardef\PSigma="7106
 \mathchardef\POmega="710A
 \mathchardef\PUpsilon="7107
                  
 \def\PB      {\ensuremath{B}\xspace}                 
                  
 \def\PD      {\ensuremath{D}\xspace}

 \def\PJ      {\ensuremath{J}\xspace}                 
 \def\PK      {\ensuremath{K}\xspace}

 \def\Pi      {\ensuremath{i}\xspace}

}

\def\mup        {\ensuremath{\Pmu^+}\xspace}
\def\mun        {\ensuremath{\Pmu^-}\xspace} 
\def\mumu       {\ensuremath{\Pmu^+\Pmu^-}\xspace}

\def\pion  {\ensuremath{\Ppi}\xspace}

\def\pip   {\ensuremath{\pion^+}\xspace}
\def\pim   {\ensuremath{\pion^-}\xspace}

\def\kaon  {\ensuremath{\PK}\xspace}
  \def\Kbar  {\kern 0.2em\overline{\kern -0.2em \PK}{}\xspace}

\def\Kz    {\ensuremath{\kaon^0}\xspace}
\def\Kzb   {\ensuremath{\Kbar^0}\xspace}
\def\KzKzb {\ensuremath{\Kz \kern -0.16em \Kzb}\xspace}
\def\Kp    {\ensuremath{\kaon^+}\xspace}
\def\Km    {\ensuremath{\kaon^-}\xspace}

\def\KpKm  {\ensuremath{\Kp \kern -0.16em \Km}\xspace}

  \def\Dbar    {\kern 0.2em\overline{\kern -0.2em \PD}{}\xspace}
\def\D       {\ensuremath{\PD}\xspace}

\def\Dz      {\ensuremath{\D^0}\xspace}
\def\Dzb     {\ensuremath{\Dbar^0}\xspace}
\def\DzDzb   {\ensuremath{\Dz {\kern -0.16em \Dzb}}\xspace}
\def\Dp      {\ensuremath{\D^+}\xspace}
\def\Dm      {\ensuremath{\D^-}\xspace}

\def\DpDm    {\ensuremath{\Dp {\kern -0.16em \Dm}}\xspace}
\def\Dstar   {\ensuremath{\D^*}\xspace}

\def\B       {\ensuremath{\PB}\xspace}
  \def\Bbar    {\kern 0.18em\overline{\kern -0.18em \PB}{}\xspace}

\def\Bu      {\ensuremath{\B^+}\xspace}

\def\Bp      {\ensuremath{\Bu}\xspace}

\def\jpsi     {\ensuremath{{\PJ\mskip -3mu/\mskip -2mu\Ppsi\mskip 2mu}}\xspace}
\def\psitwos  {\ensuremath{\Ppsi{(2S)}}\xspace}

  \def\Y#1S{\ensuremath{\PUpsilon{(#1S)}}\xspace}

\newcommand{\decay}[2]{\ensuremath{#1\!\to #2}\xspace}         

\def\to                 {\ensuremath{\rightarrow}\xspace}

\def\AT#1     {\ensuremath{A_T^{#1}}\xspace}           

\def\C#1      {\ensuremath{\mathcal{C}_{#1}}\xspace}                       
\def\Cp#1     {\ensuremath{\mathcal{C}_{#1}^{'}}\xspace}                    
\def\Ceff#1   {\ensuremath{\mathcal{C}_{#1}^{\mathrm{(eff)}}}\xspace}        
\def\Cpeff#1  {\ensuremath{\mathcal{C}_{#1}^{'\mathrm{(eff)}}}\xspace}       
\def\Ope#1    {\ensuremath{\mathcal{O}_{#1}}\xspace}                       
\def\Opep#1   {\ensuremath{\mathcal{O}_{#1}^{'}}\xspace}                    

\newcommand{\tev}{\ensuremath{\mathrm{\,Te\kern -0.1em V}}\xspace}
\newcommand{\gev}{\ensuremath{\mathrm{\,Ge\kern -0.1em V}}\xspace}
\newcommand{\mev}{\ensuremath{\mathrm{\,Me\kern -0.1em V}}\xspace}
\newcommand{\kev}{\ensuremath{\mathrm{\,ke\kern -0.1em V}}\xspace}
\newcommand{\ev}{\ensuremath{\mathrm{\,e\kern -0.1em V}}\xspace}
\newcommand{\gevc}{\ensuremath{{\mathrm{\,Ge\kern -0.1em V\!/}c}}\xspace}
\newcommand{\mevc}{\ensuremath{{\mathrm{\,Me\kern -0.1em V\!/}c}}\xspace}
\newcommand{\gevcc}{\ensuremath{{\mathrm{\,Ge\kern -0.1em V\!/}c^2}}\xspace}
\newcommand{\gevgevcccc}{\ensuremath{{\mathrm{\,Ge\kern -0.1em V^2\!/}c^4}}\xspace}
\newcommand{\mevcc}{\ensuremath{{\mathrm{\,Me\kern -0.1em V\!/}c^2}}\xspace}

\def\invpb {\ensuremath{\mbox{\,pb}^{-1}}\xspace}

\newcommand{\stat}{\ensuremath{\mathrm{(stat)}}\xspace}
\newcommand{\syst}{\ensuremath{\mathrm{(syst)}}\xspace}

\newcommand{\chisq}{\ensuremath{\chi^2}\xspace}

\def\gsim{{~\raise.15em\hbox{$>$}\kern-.85em
          \lower.35em\hbox{$\sim$}~}\xspace}
\def\lsim{{~\raise.15em\hbox{$<$}\kern-.85em
          \lower.35em\hbox{$\sim$}~}\xspace}

\def\sqs   {\ensuremath{\protect\sqrt{s}}\xspace}

\def\pt         {\mbox{$p_{\rm T}$}\xspace}

\def\dllkpi     {\ensuremath{\mathrm{DLL}_{\kaon\pion}}\xspace}

\def\mrad{\ensuremath{\rm \,mrad}\xspace}

\def\pythia     {\mbox{\textsc{Pythia}}\xspace}

\def\geant      {\mbox{\textsc{Geant4}}\xspace}

\def\tell1  {TELL1\xspace}
\def\ukl1   {UKL1\xspace}

\def\BpKMuMu     {\decay{\Bp}{\Kp \mup\mun}}
\def\BpKMuMuSS   {\decay{\Bp}{\Km \mup\mup}}
\def\BphMuMu     {\decay{\Bp}{ h^{+} \mup\mun}}
\def\BphMuMuSS   {\decay{\Bp}{ h^{-} \mup\mup}}
\def\BpPiMuMu     {\decay{\Bp}{\pip \mup\mun}}
\def\BpPiMuMuSS   {\decay{\Bp}{\pim \mup\mup}}

\def\BphMuMuBoth {\decay{\Bp}{h^\pm\mup\mu^\mp}}

\def\mutoK {\decay{\mu}{K}}

\def\BpJpsiK     {\decay{\Bp}{\jpsi\Kp}}
\def\BpJpsiPi     {\decay{\Bp}{\jpsi\pip}}

\begin{document}
\title{Search for the lepton number violating decays $\boldsymbol{B^{+}
    \rightarrow \pi^- \mu^+ \mu^+}$ and $\boldsymbol{B^{+}
    \rightarrow K^- \mu^+ \mu^+}$}
\author{
R.~Aaij$^{23}$, 
C.~Abellan~Beteta$^{35,n}$, 
B.~Adeva$^{36}$, 
M.~Adinolfi$^{42}$, 
C.~Adrover$^{6}$, 
A.~Affolder$^{48}$, 
Z.~Ajaltouni$^{5}$, 
J.~Albrecht$^{37}$, 
F.~Alessio$^{37}$, 
M.~Alexander$^{47}$
% ...

{\it \footnotesize
% ...
$ ^{5}$Clermont Universit\'{e}, Universit\'{e} Blaise Pascal, CNRS/IN2P3, LPC, Clermont-Ferrand, France\\
$ ^{6}$CPPM, Aix-Marseille Universit\'{e}, CNRS/IN2P3, Marseille, France\\
% ...
$ ^{23}$Nikhef National Institute for Subatomic Physics, Amsterdam, Netherlands\\
% ...
$ ^{35}$Universitat de Barcelona, Barcelona, Spain\\
$ ^{36}$Universidad de Santiago de Compostela, Santiago de Compostela, Spain\\
$ ^{37}$European Organization for Nuclear Research (CERN), Geneva, Switzerland\\
% ...
$ ^{42}$H.H. Wills Physics Laboratory, University of Bristol, Bristol, United Kingdom\\
% ...
$ ^{47}$School of Physics and Astronomy, University of Glasgow, Glasgow, United Kingdom\\
$ ^{48}$Oliver Lodge Laboratory, University of Liverpool, Liverpool, United Kingdom\\
% ...
$ ^{n}$LIFAELS, La Salle, Universitat Ramon Llull, Barcelona, Spain\\
% ...
}
}
\begin{abstract}
  \vspace{0.1cm}
  \noindent A search is performed for the lepton number violating
  decay \BphMuMuSS, where $h^-$ represents a \Km or a \pim, using data from
  the LHCb detector corresponding to an integrated luminosity of
  $36\mbox{\,pb}^{-1}$. The decay is forbidden in the
  Standard Model but allowed in models with a Majorana
  neutrino. No signal is observed in either channel and limits of
  ${\cal B}(B^{+} \rightarrow K^- \mu^+ \mu^+) < 5.4
  \times 10^{-8}$ and ${\cal B}(B^{+} \rightarrow \pi^- \mu^+ \mu^+) < 5.8
  \times 10^{-8}$ are set at the 95\% confidence level. These improve
  the previous best limits by factors of 40 and 30, respectively.
\end{abstract}

\pacs{11.30.Fs, 13.20.He, 13.35.Hb}

\vspace*{-1cm}
\hspace{-9cm}
\mbox{\Large EUROPEAN ORGANIZATION FOR NUCLEAR RESEARCH (CERN)}

\vspace*{0.2cm}
\hspace*{-9cm}
\begin{tabular*}{16cm}{lc@{\extracolsep{\fill}}r}
\ifthenelse{\boolean{pdflatex}}
{\vspace*{-2.7cm}\mbox{\!\!\!\includegraphics[width=.14\textwidth]{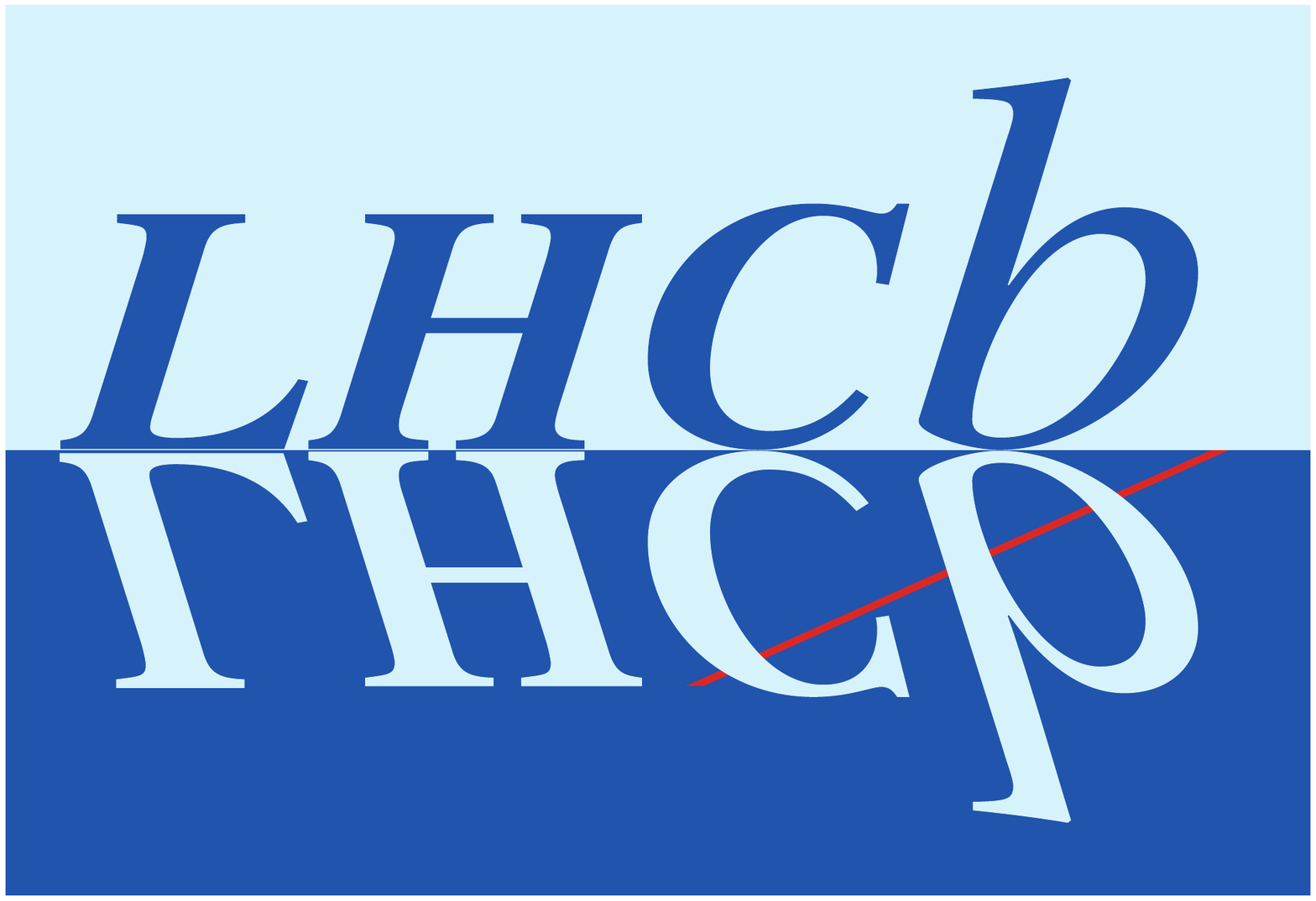}} & &}%
{\vspace*{-1.2cm}\mbox{\!\!\!\includegraphics[width=.12\textwidth]{figs/lhcb-logo.eps}} & &}%
\\
 & & LHCb-PAPER-2011-009 \\
 & & CERN-PH-EP-2011-156 \\ 
 & & \today \\ 
\end{tabular*}
\vspace*{1cm}

\maketitle

\begin{figure}[b]
 \begin{center}
 \includegraphics[width=.45\textwidth]{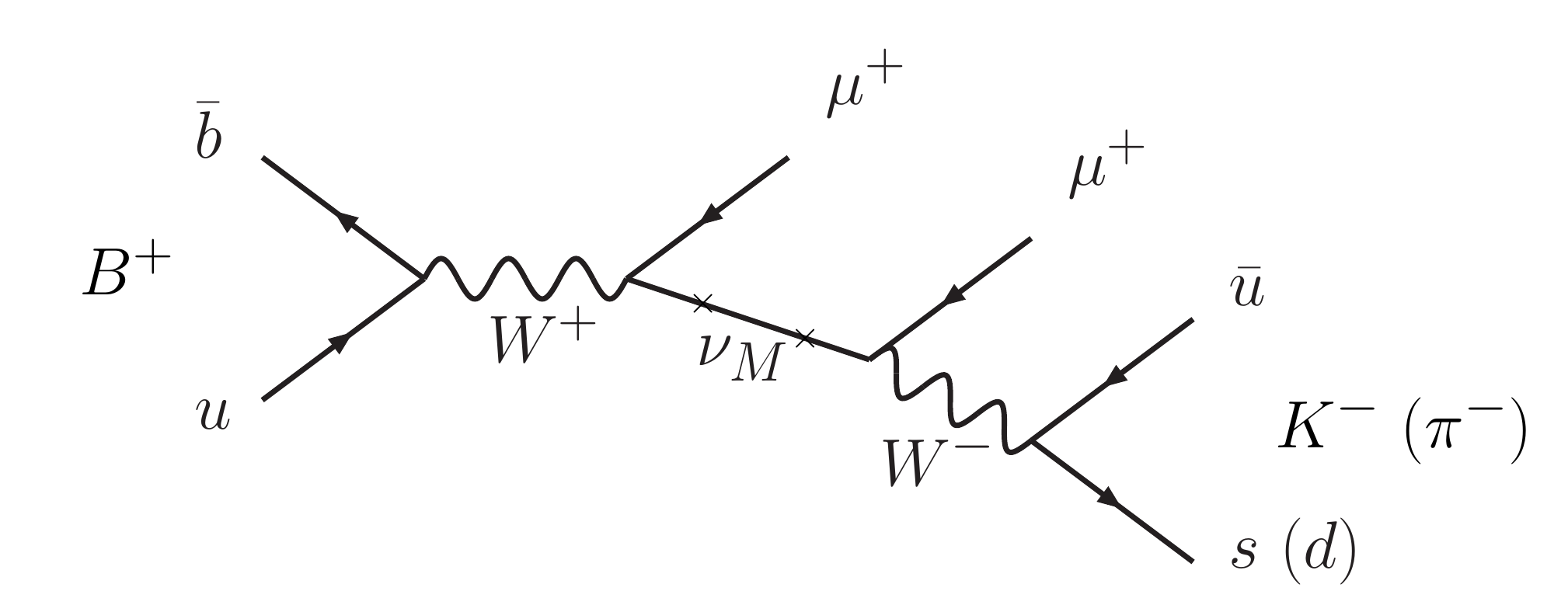}
 \caption{s-channel diagram for \BpKMuMuSS \mbox{(\BpPiMuMuSS)} where the decay is mediated by an on-shell Majorana neutrino.}
 \label{fig:onshelldiagram}
\end{center}
\end{figure}

\noindent Lepton number is conserved in the Standard Model but
can be violated in a range of new physics models such as those with Majorana
neutrinos~\cite{Majorana:1937vz} or left-right symmetric models with a
doubly charged Higgs boson~\cite{Pati:1974yy}. In this letter a search
for lepton number violating decays of the type \BphMuMuSS, where $h^-$
represents a \Km or a \pim, is presented. The inclusion of charge
conjugated modes is implied throughout. 
A search for any lepton number violating process that mediates the \BphMuMuSS decay is made.
A specific search for \BphMuMuSS decays mediated by an on-shell Majorana neutrino (Fig.~\ref{fig:onshelldiagram}) is also performed.
Such decays would 
give rise to a
narrow peak in the invariant mass spectrum of the hadron and one of the
muons~\cite{Pascoli:2007qh}, $m_\nu=m_{h\mu}$, if the mass of the
neutrino is between $m_{K(\pi)}+m_\mu$ and $m_\B - m_\mu$. The previous
best experimental limit on the \decay{\Bp}{\Km(\pim)\mup\mup}
branching fraction is $1.8 (1.2) \times 10^{-6}$ at 90\% confidence
level~(CL)~\cite{Edwards:2002kq}.

The search for \BphMuMuSS is carried out with data from the \lhcb
experiment at the Large Hadron Collider at CERN. The data corresponds
to 36\invpb of integrated luminosity of proton-proton collisions at
$\sqs=7\tev$ collected in 2010. The \lhcb detector is a single-arm spectrometer designed
to study $b$-hadron decays with an acceptance for charged tracks with
pseudorapidity between 2 and 5. Primary proton-proton vertices~(PVs),
and secondary \B vertices are identified in a silicon strip vertex
detector. Tracks from charged particles are reconstructed by the
vertex detector and a set of tracking stations. The curvature of the
tracks in a dipole magnetic field allows momenta to be determined with a
precision of $\delta p/p =0.35$--$0.5\%$. Two Ring Imaging CHerenkov
(RICH) detectors allow kaons to be separated from pions/muons over a
momentum range $2<p<100\gevc$. Muons with momentum above 3\gevc are
identified on the basis of the number of hits left in detectors
interleaved with an iron muon filter. Further details about the \lhcb
detector can be found in Ref.~\cite{Alves:2008zz}.

The search for \BphMuMuSS decays is based on the selection of
\BphMuMuBoth candidates. The \BpJpsiK decay with
\decay{\jpsi}{\mup\mun} is included in the same selection. It is
subsequently used as a normalisation mode when setting a limit on the
branching fraction of the \BphMuMuSS decays.  The selection is
designed to minimise and control the difference between decays with
same- and opposite-sign muons and thus cancel most of the
systematic uncertainty from the normalisation. The only differences in
efficiency between the signal and normalisation channels are due to
the decay kinematics and the presence of a same-sign muon pair, rather
than an opposite-sign pair, in the final state.

In the trigger, the \BphMuMuBoth candidates are required to pass the
initial hardware trigger based on the \pt of one of the muons. In the
subsequent software trigger, one of the muons is required to have a
large impact parameter~(IP) with respect to all the PVs in the event
and to pass requirements on the quality of the track fit and the
compatibility of the candidate with the muon hypothesis. Finally, the
muon candidate combined with another track is required to form a
vertex displaced from the PVs.

Further event selection is applied offline on fully reconstructed \B
decay candidates. The selection is designed to reduce combinatorial backgrounds,
where not all the selected tracks come from the same decay vertex; and
peaking backgrounds, where a single decay is selected but with some of
the particle types misidentified.  The combinatorial background is
smoothly distributed in the reconstructed \B-candidate mass and the
level of background is assessed from the sidebands around the signal
window. Peaking backgrounds from \B decays to hadronic final states,
final states with a \jpsi and semileptonic final states are also considered.

Proxies are used in the optimisation of the selection for both the
signal and the background to avoid a selection bias. The \BpJpsiK decay is used as a proxy for the signal. The background proxy comprises opposite-sign
\BphMuMu candidates with an invariant mass in the upper mass sideband
and with muon pairs incompatible with a \jpsi or a \psitwos
hypothesis. The bias introduced by using \BpJpsiK for both optimisation and
as a normalisation mode is insignificant due to the large
number of candidates.

The combinatorial background is reduced by requiring that the decay products of the \B have $\pt > 800\mevc$. Tracks are selected which are incompatible with originating from a PV based on the \chisq of the tracks' impact parameters ($\chi^{2}_{\rm IP} > 45$). The direction of the
candidate \Bp momentum is required to be within 8\mrad of the reconstructed
\Bp line of flight. The \Bp vertex is also required to be of good
quality ($\chi^{2} < 12$ for three degrees of freedom) and significantly displaced from the PV ($\chi^{2}$ of vertex separation larger than 144).

The selection uses a range of particle identification~(PID) criteria,
based on information from the RICH and muon detectors, to ensure the
hadron and the muons are correctly identified. For example, \dllkpi is
the difference in log-likelihoods between the $\kaon$ and $\pi$ hypotheses. For
the \BpKMuMuSS final state, $\dllkpi >1$ is required to select kaon
candidates. For the \BpPiMuMuSS final state the selection criterion is
mirrored to select pions with $\dllkpi < -1$. The \BpKMuMuSS and
\BpPiMuMuSS selections are otherwise identical.

To reject background events where two tracks that are close together
in the tracking system share hits in the muon detector, a requirement is
made on the maximum number of muon system hits that two candidate
muons may have in common. This requirement can introduce a bias in the
relative efficiency between signal and normalisation channels, as both
tracks from the same-sign muon pair for \BphMuMuSS will curve in the
same direction in the dipole field. Simulated events give an estimate of 0.3\% for the effect on the relative efficiency between the signal and
normalisation channel. In order to avoid selecting a muon as the pion or kaon, the candidate
hadron is also required to be within the acceptance of the muon system
but not have a track segment there.

After the application of the above criteria the combinatorial
background is completely dominated by candidates with two real muons,
rather than by hadrons mis-identified as muons.

The invariant mass distribution and the relevant mis-identification
rates are required in order to evaluate the peaking background. These
are evaluated, respectively, from a full simulation using
\pythia~\cite{Sjostrand:2006za} followed by
\geant~\cite{Agostinelli:2002hh}, and from control channels which
provide an unambiguous and pure source of particles of known type. The
control channel events are selected to have the same kinematics as the
signal decay, without the application of any PID criteria.
$\decay{\Dstar}{\Dz \pi}$, $\decay{\Dz}{K\pi}$ decays give pure
sources of pions and kaons. A pure source of muons is selected by
using a \textit{tag-and-probe} approach with
\decay{\jpsi}{\mumu} decays~\cite{Aaij:2011rj}.

Under the \BpKMuMuSS hypothesis, any crossfeed from \BpJpsiK decays
would peak strongly in the signal mass region. The $\kaon\to\mu$ mis-id rate is evaluated from the above \Dstar sample and the \mutoK mis-id rate from the \jpsi sample. The later mis-id rate is consistent with zero.  The number of \BpJpsiK events expected in the signal
region is therefore also zero but with a large uncertainty which
dominates the error on the total exclusive background expected in the
signal region. The $\decay{\Bp}{\pim\pip\Kp}$ decay contributes the
most to the peaking background with an expected $(1.7 \pm 0.1) \times
10^{-3}$ candidates, followed by the $\decay{\Bp}{\Km\pip\Kp}$ decay
with $(6.1 \pm 0.8) \times 10^{-4}$ candidates. The total peaking
background expected in the \BpKMuMuSS signal region is
$(3.4^{+14.0}_{-0.2}) \times 10^{-3}$ events with the asymmetric error
caused by the zero expectation from the \BpJpsiK decay.

Under the \BpPiMuMuSS hypothesis, \BpJpsiK decays are
 reconstructed with invariant masses below the nominal \Bp mass, in the
lower mass sideband. The dominant background decay in this case is $\Bp
\rightarrow \pim\pip\pip$, where the two same-sign pions are
misidentified as muons. The \BpPiMuMuSS peaking background level is
$(2.9 \pm 0.6) \times 10^{-2}$ events.

\begin{figure}[b]
  \begin{center}
    \includegraphics[width=.48\textwidth]{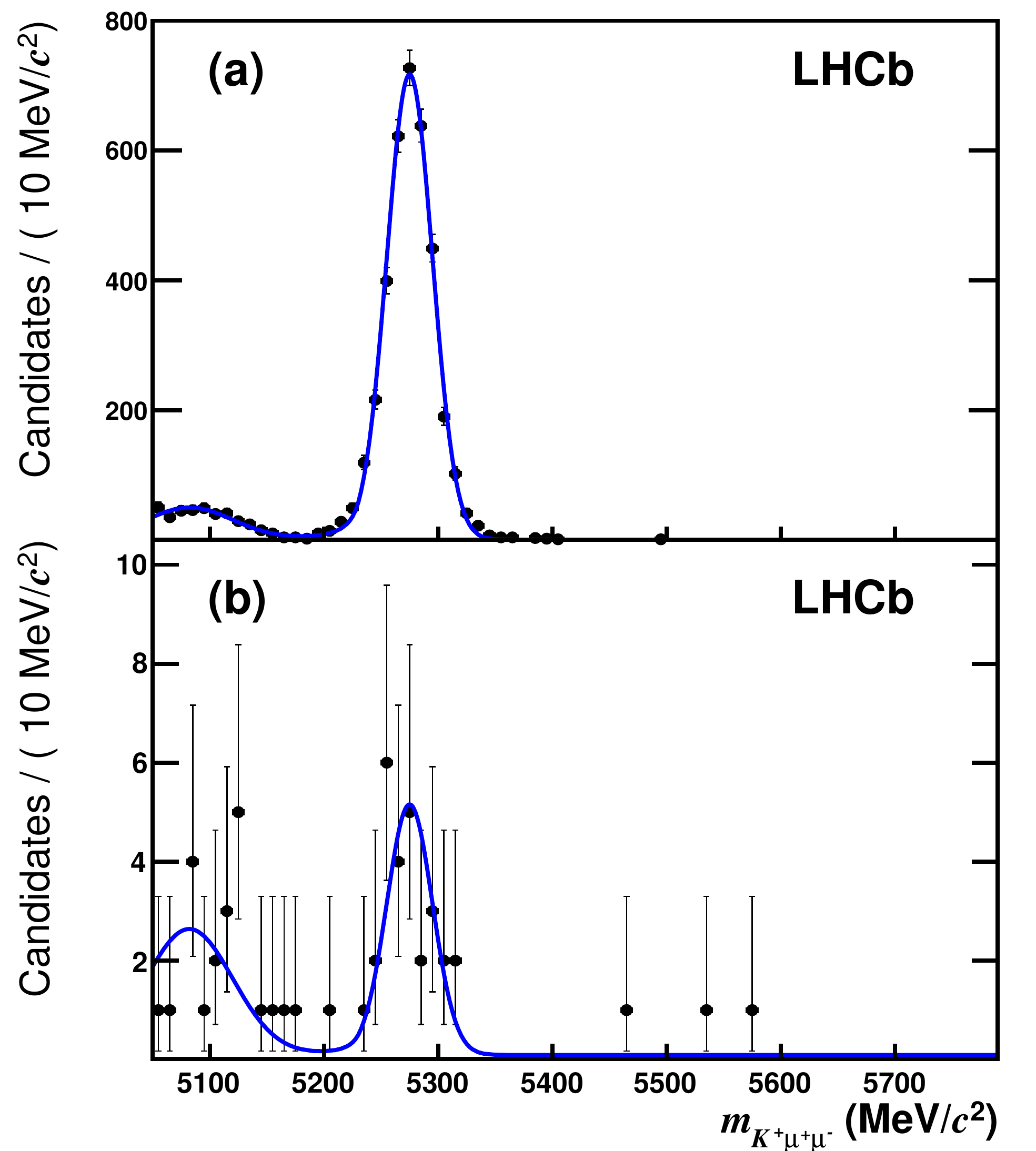}
    \caption{Invariant mass distribution of $\Kp\mup\mun$ events after
      the application of the selection criteria. In (a) requiring the
      muon pair to be compatible with coming from a \jpsi decay and in
      (b) excluding invariant mass windows around the \jpsi and
      \psitwos for the muon pair. The curve is the fit to data as
      described in the text.}
    \label{fig:kmumuos}
  \end{center}
\end{figure}

In Fig.~\ref{fig:kmumuos}(a), the $m_{\Kp\mup\mun}$ invariant mass
distribution for \BpKMuMu events with
\mbox{$|m_{\mup\mun}-m_{\jpsi}|<50\mevcc$} is shown, after the
application of the selection. In the \BpJpsiK sample, there are no events containing more than one candidate. An unbinned maximum likelihood fit to the
\BpJpsiK mass peak is made with a Crystal
Ball~\cite{Skwarnicki:1986xj} function which accounts for the
radiative tail. The combinatorial background is assumed to be flat,
and the partially reconstructed events in the lower mass sideband are
fitted with a Gaussian distribution.  The signal peak has a Gaussian component of
width 20\mevcc, and a signal mass window of $5280 \pm 40\mevcc$ is
chosen.  The \BpJpsiK peak contains $3407 \pm 59$ signal
events within the signal window.  \BpJpsiPi candidates were also
examined and, accounting for a shoulder in the mass distribution from
\BpJpsiK, the yield observed agrees with the expectation.

The $m_{\Kp\mup\mun}$ invariant mass distribution for events with
\mbox{$|m_{\mup\mun}-m_{\jpsi,\psitwos}|>70\mevcc$} is shown in
Fig.~\ref{fig:kmumuos}(b).  Using the same fit model, with all
shape parameters fixed to those from the above fit, the signal peak
was determined to contain $27\pm 5$ events from the \BpKMuMu decay.
The ratio of branching fractions between \BpJpsiK and \BpKMuMu
decays~\cite{Nakamura:2010zzi} and the trigger efficiency ratio predicted by the
simulation, give an expectation of $29\pm 4$ \BpKMuMu decays. The
observed yield is consistent with the expectation showing that the
selection does not favour candidates with a
dimuon mass close to the \jpsi mass.

\begin{figure}[b]
 \begin{center}
 \vspace{-0.5cm}
 \includegraphics[width=0.45\textwidth]{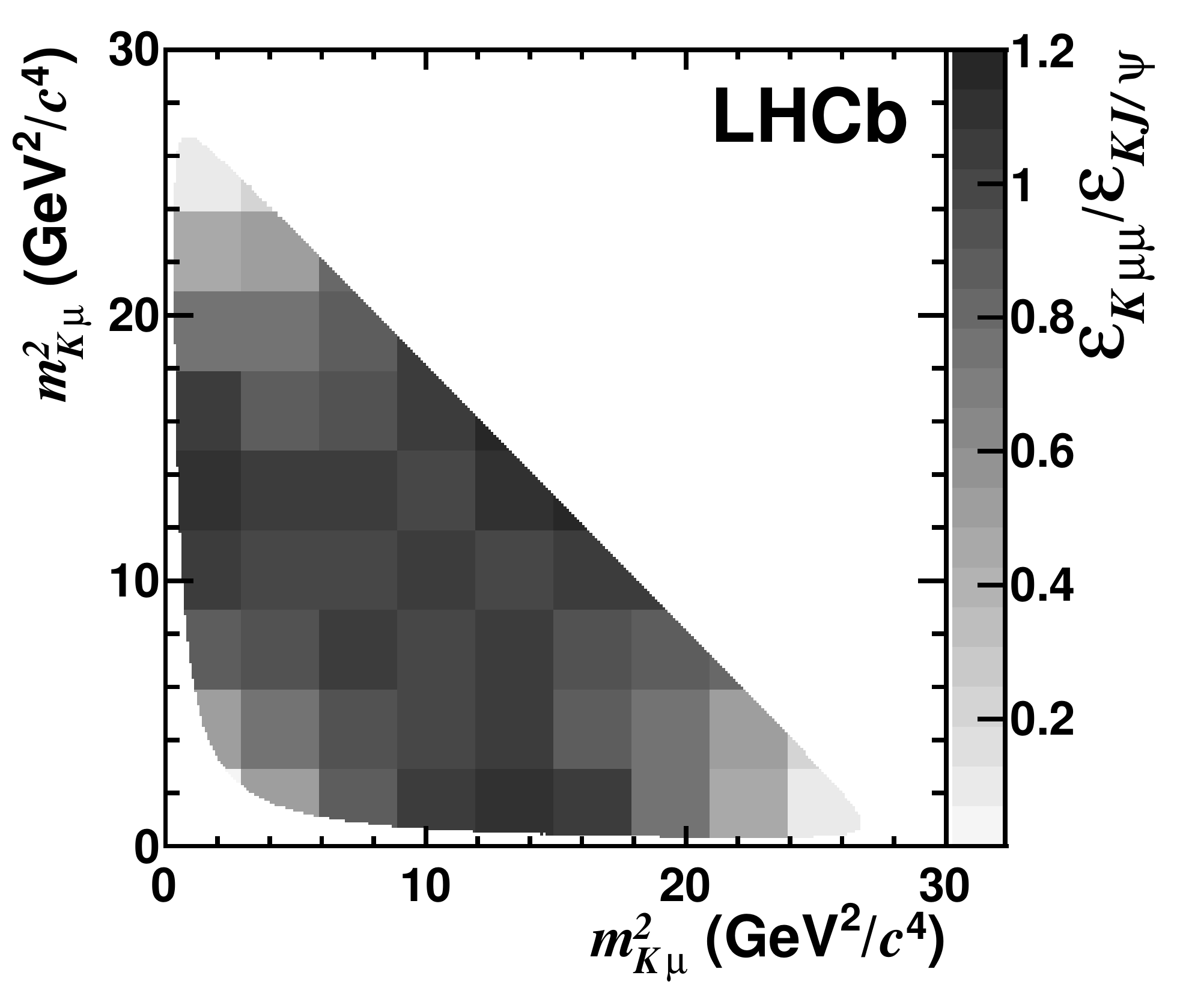}
 \vspace{-0.5cm}
 \caption{Relative efficiency between the \BpKMuMuSS signal and the
   \BpJpsiK normalisation channel. The plot has been symmetrised over
   the diagonal.}
 \label{fig:efficiencyexample}
\end{center}
\end{figure}

The difference in efficiency between the signal and normalisation
channels was evaluated using Monte Carlo simulation samples. The
relative selection efficiency across the phase space is shown for
\BpKMuMuSS in Fig.~\ref{fig:efficiencyexample}. The efficiency of the
signal selection in a given phase space bin is divided by the average
efficiency of \BpJpsiK, to yield the relative efficiency for that bin.
The \Dstar control channel is used to determine the PID efficiencies
required to normalise \BpPiMuMuSS to \BpJpsiK.

Assuming a signal that is uniformly distributed in phase space,
 the relative efficiency of
\BpKMuMuSS and \BpJpsiK was calculated to be $89.1 \pm
0.4\,\stat\pm 0.3\,\syst \%$. The relative
efficiency of \BpPiMuMuSS and \BpJpsiK was calculated to be $82.7 \pm
0.6\,\stat\pm 0.8\,\syst\%$. The systematic
uncertainties associated with these estimates are detailed below.
These relative efficiencies together with the number of events
observed in the normalisation channel, give single event sensitivities
of $2.0 \times 10^{-8}$ ($2.1 \times 10^{-8}$) in the \BpKMuMuSS
(\BpPiMuMuSS) case.

In order to compute the efficiency under a given Majorana neutrino
mass hypothesis, a model for the variation of efficiency with
$m_{h\mu}$ is required. For a given value of $m_{h\mu}$ this is obtained
by varying the polarisation of the Majorana neutrino in the decay and
taking the lowest (most conservative) value of the efficiency.

The dominant systematic uncertainty (under the assumption of a flat phase-space distribution) for the single event sensitivity
is the 3.4\% uncertainty on the \BpJpsiK branching fraction. The
statistical uncertainty on the \BpJpsiK yield gives an additional
systematic uncertainty of 1.7\% and the uncertainty from the model
used to fit the data is 1.6\%. 
The latter is evaluated by changing the
Crystal Ball signal function used in the
fit to a Gaussian and the polynomial background function to an exponential.

There are several sources of uncertainty associated with the
calculation of the relative efficiency between the signal and
normalisation channels. In addition to the statistical uncertainty of
the simulation samples, there are systematic uncertainties from: the
differences in the effect of the IP selection criteria between the
simulation and data; the statistical uncertainty on the measured PID
efficiencies; the uncertainties associated with the simulation of the
trigger; and the uncertainty in the tracking
efficiency. In each case the systematic uncertainty is estimated by
varying the relevant criteria at the level of the expected effect and
re-evaluating the relative efficiency.  For the \BpPiMuMuSS decay,
there is an additional uncertainty from the correction for the
relative kaon- and pion-identification efficiencies. The systematic
uncertainties averaged over the three-body phase space are given in
Table~\ref{table:sys:eff}.

\begin{table}
 \caption{
    Sources of systematic error and their fractional uncertainty on the 
relative efficiency. 
    \label{table:sys:eff}
  }
  \centering
  \begin{tabular}{|l|c|c|}
    \hline
    Source & \BpKMuMuSS  & \BpPiMuMuSS \\
    \hline 
    \hline
    $\mathcal{B}(\BpJpsiK)$ & 3.4\% & 3.4\% \\
    \BpJpsiK yield & 1.7\% & 1.7\%\\
    \BpJpsiK fit models & 1.6\% & 1.6\% \\
    Simulation statistics & 0.4\% & 0.6\%\\
    IP modelling & 0.2\% & 0.2\%\\ 
    PID modelling & 0.1\% & 0.8\%\\
    Trigger efficiency & 0.1\% & 0.1\%\\
    Tracking efficiency & 0.1\% & 0.1\% \\
    \hline 
  \end{tabular}
 \end{table}

A limit on the branching fraction of each of the \BphMuMuSS decays is
set by counting the number of observed events in
the mass windows, and using the single event sensitivity. The probability is
modelled with a Poisson distribution where the mean has contributions
from a potential signal, the combinatorial and peaking backgrounds.
The combinatorial background is unconstrained by measurements from the
simulation or the opposite-sign data. The number of events in the
upper mass sideband is therefore used to constrain the contribution
of the combinatorial background to the Poisson mean. The upper mass
sideband is restricted to masses above $m_{h\mu\mu} > 5.4\gevcc$ such
that any peaking background component can be ignored. In both the
\BpKMuMuSS and \BpPiMuMuSS cases no events are found in either the
upper or lower mass sidebands. This is consistent with the observation of three opposite-sign candidates seen in the \BpKMuMu upper mass sideband (Fig.~\ref{fig:kmumuos}) and two candidates in the \BpPiMuMu upper mass sideband. The peaking background estimates are
explicitly split into two components, the contribution from $\Bp \to
h^{-} h^{+} h^{+}$ decays and that from \BpJpsiK decays. The latter
has a large uncertainty. The central values for both peaking
background components are taken from the estimates described above.

Systematic uncertainties on the peaking background, single event
sensitivity and signal-to-sideband scale factor are included in the
limit-setting procedure using a Bayesian approach. The unknown parameter is
integrated over and included in the probability to observe a given
number of events in the signal and upper mass window.

In the signal mass windows of \BpKMuMuSS and \BpPiMuMuSS no events are
observed. This corresponds to limits on the \BphMuMuSS branching
fractions of
\begin{displaymath}
\begin{split}
 \mathcal{B}(\BpKMuMuSS) & < 5.4\,(4.1) \times 10^{-8} ~\text{at 95\%\,(90\%) CL}, \\
 \mathcal{B}(\BpPiMuMuSS) & < 5.8\,(4.4) \times 10^{-8} ~\text{at 95\%\,(90\%) CL}. \\
\end{split}
\end{displaymath}

\begin{figure}[htb]
  \begin{center}
    \includegraphics[width=0.45\textwidth]{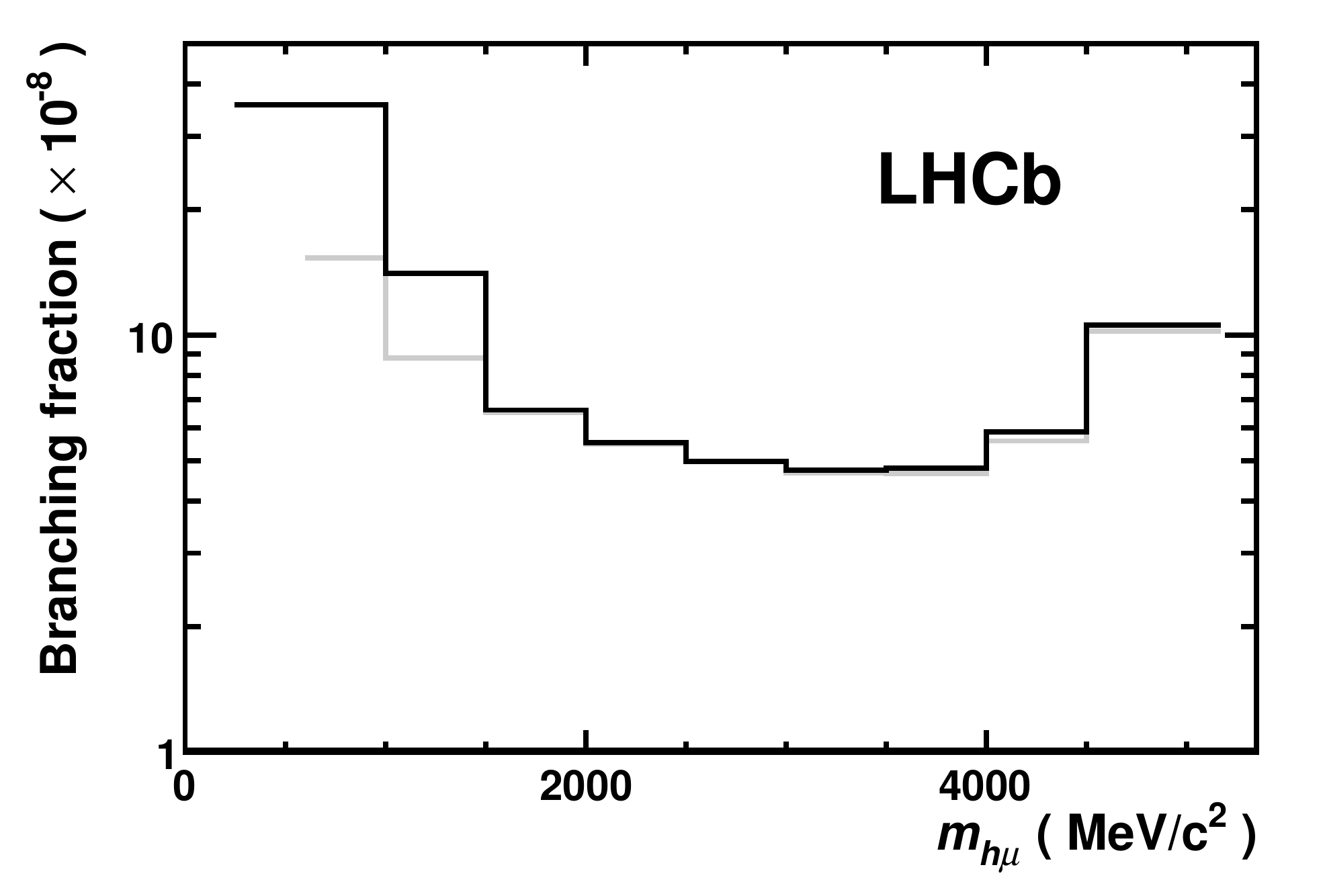}
    \caption{The 95 \% CL branching fraction limits for \BpKMuMuSS
      (light-coloured line) and \BpPiMuMuSS (dark-coloured line) as a
      function of the Majorana neutrino mass $m_\nu=m_{h\mu}$.}
    \label{fig:observed:mass}
  \end{center}
\end{figure} 

The observation of no candidates in the sidebands as well as the
signal region is compatible with a background-only
hypothesis. The $m_{h\mu}$ dependence of the limit in models where the
Majorana neutrino can be produced on mass shell is shown in
Fig.~\ref{fig:observed:mass}. The shapes of the limits arise from the
changing efficiency as a function of mass.

In summary, a search for the \BpKMuMuSS and \BpPiMuMuSS decays modes
has been performed with $36\mbox{\,pb}^{-1}$ of integrated luminosity
collected with the LHCb detector in 2010. No signal is observed in
either decay and, using \BpJpsiK as a normalisation channel, the
present best limits on ${\cal B}(B^{+} \rightarrow K^- \mu^+ \mu^+)$
and ${\cal B}(B^{+} \rightarrow \pi^- \mu^+ \mu^+)$ are improved by
factors of 40 and 30, respectively~\cite{Edwards:2002kq}.

\section*{Acknowledgments}

\noindent We express our gratitude to our colleagues in the CERN accelerator
departments for the excellent performance of the LHC. We thank the
technical and administrative staff at CERN and at the LHCb institutes,
and acknowledge support from the National Agencies: CAPES, CNPq,
FAPERJ and FINEP (Brazil); CERN; NSFC (China); CNRS/IN2P3 (France);
BMBF, DFG, HGF and MPG (Germany); SFI (Ireland); INFN (Italy); FOM and
NWO (Netherlands); SCSR (Poland); ANCS (Romania); MinES of Russia and
Rosatom (Russia); MICINN, XuntaGal and GENCAT (Spain); SNSF and SER
(Switzerland); NAS Ukraine (Ukraine); STFC (United Kingdom); NSF
(USA). We also acknowledge the support received from the ERC under FP7
and the R\'egion Auvergne.

\bibliographystyle{LHCb}
\bibliography{main}

\providecommand{\href}[2]{#2}\begingroup\raggedright\begin{thebibliography}{10}

\bibitem{Majorana:1937vz}
E.~Majorana, {\it {Teoria simmetrica dell'elettrone e del positrone}},
  \href{http://dx.doi.org/10.1007/BF02961314}{ {\em Nuovo Cim.} {\bf 14} (1937)
  171--184}.

\bibitem{Pati:1974yy}
J.~C. Pati and A.~Salam, {\it {Lepton Number as the Fourth Color}},
  \href{http://dx.doi.org/10.1103/PhysRevD.10.275}{ {\em Phys. Rev.} {\bf D10}
  (1974) 275--289}. Erratum-ibid. {\bf D11} (1975) 703.

\bibitem{Pascoli:2007qh}
S.~Pascoli and S.~Petcov, {\it {Majorana Neutrinos, Neutrino Mass Spectrum and
  the $|\langle m \rangle| \sim 10^{-3}$~eV Frontier in Neutrinoless Double
  Beta Decay}},  \href{http://dx.doi.org/10.1103/PhysRevD.77.113003}{ {\em
  Phys. Rev.} {\bf D77} (2008) 113003},
  [\href{http://xxx.lanl.gov/abs/0711.4993}{{\tt arXiv:0711.4993}}].

\bibitem{Edwards:2002kq}
CLEO collaboration, K.~W. Edwards et~al., {\it {Search for lepton flavor
  violating decays of B mesons}},
  \href{http://dx.doi.org/10.1103/PhysRevD.65.111102}{ {\em Phys. Rev.} {\bf
  D65} (2002) 111102}, [\href{http://xxx.lanl.gov/abs/hep-ex/0204017}{{\tt
  arXiv:hep-ex/0204017}}].

\bibitem{Alves:2008zz}
LHCb collaboration, A.~A. Alves et~al., {\it {The LHCb Detector at the LHC}},
  \href{http://dx.doi.org/10.1088/1748-0221/3/08/S08005}{ {\em JINST} {\bf 3}
  (2008) S08005}.

\bibitem{Sjostrand:2006za}
T.~{Sj\"{o}strand}, S.~Mrenna, and P.~Z. Skands, {\it {PYTHIA 6.4 Physics and
  Manual}},  \href{http://dx.doi.org/10.1088/1126-6708/2006/05/026}{ {\em JHEP}
  {\bf 05} (2006) 026}, [\href{http://xxx.lanl.gov/abs/hep-ph/0603175}{{\tt
  arXiv:hep-ph/0603175}}].

\bibitem{Agostinelli:2002hh}
GEANT4 collaboration, S.~Agostinelli et~al., {\it {GEANT4: A simulation
  toolkit}},  \href{http://dx.doi.org/10.1016/S0168-9002(03)01368-8}{ {\em
  Nucl. Instrum. Meth.} {\bf A506} (2003) 250--303}.

\bibitem{Aaij:2011rj}
LHCb collaboration, R.~Aaij et~al., {\it {Search for the rare decays $\Bs \to
  \mumu$ and $\Bd \to \mumu$}},
  \href{http://dx.doi.org/10.1016/j.physletb.2011.04.031}{ {\em Phys. Lett.}
  {\bf B699} (2011) 330--340}, [\href{http://xxx.lanl.gov/abs/1103.2465}{{\tt
  arXiv:1103.2465}}].

\bibitem{Skwarnicki:1986xj}
T.~Skwarnicki, {\em {A study of the radiative cascade transitions between the
  Upsilon-prime and Upsilon resonances}}.
\newblock PhD thesis, Institute of Nuclear Physics, Krakow, 1986.
\newblock DESY-F31-86-02.

\bibitem{Nakamura:2010zzi}
Particle Data Group, K.~Nakamura et~al., {\it {Review of particle physics}},
  \href{http://dx.doi.org/10.1088/0954-3899/37/7A/075021}{ {\em J. Phys.} {\bf
  G37} (2010) 075021}.

\end{thebibliography}\endgroup

\end{document}